\documentclass[prl,aps,showpacs,twocolumn,superscriptaddress]{revtex4}
\usepackage{graphicx,bm}
\usepackage{amssymb}
\usepackage{epsfig}
\usepackage{epsf}
\usepackage[usenames]{color}

\begin{document}

\title{Spin-vorticity coupling in viscous electron fluids}

\author{Ruben J. Doornenbal}

\affiliation{Institute for Theoretical Physics, Utrecht
	University, Leuvenlaan 4, 3584 CE Utrecht, The Netherlands}

\author{Marco Polini}

\affiliation{Istituto Italiano di Tecnologia, Graphene Labs, Via Morego 30, I-16163 Genova, Italy}

\author{Rembert A. Duine}

\affiliation{Institute for Theoretical Physics, Utrecht
University, Leuvenlaan 4, 3584 CE Utrecht, The Netherlands}

\affiliation{Department of Applied Physics, Eindhoven University of Technology, P.O. Box 513, 5600 MB Eindhoven, The Netherlands}

\date{\today}

\begin{abstract}
We consider spin-vorticity coupling---the generation of spin polarization by vorticity---in viscous two-dimensional electron systems with spin-orbit coupling. We first derive hydrodynamic equations for spin and momentum densities in which their mutual coupling is determined by the rotational viscosity. We then calculate the rotational viscosity microscopically in the limits of weak and strong spin-orbit coupling. We provide estimates that show that the spin-orbit coupling achieved in recent experiments is strong enough for the spin-vorticity coupling to be observed. On the one hand, this coupling provides a way to image viscous electron flows by imaging spin densities. On the other hand, we show that the spin polarization generated by spin-vorticity coupling in the hydrodynamic regime can, in principle, be much larger than that generated, e.g. by the spin Hall effect, in the diffusive regime. 
\end{abstract}


\maketitle

% definitions
\def\bx{{\bm x}}
\def\bk{{\bm k}}
\def\bK{{\bm K}}
\def\bq{{\bm q}}
\def\br{{\bm r}}
\def\bp{{\bm p}}
\def\bM{{\bm M}}
\def\bs{{\bm s}}
\def\bB{{\bm B}}
\def\bj{{\bm j}}
\def\bF{{\bm F}}
\def\id{{\rm d}}

\def\br{{\bm r}}
\def\bv{{\bm v}}

\def\half{\frac{1}{2}}
\def\args{(\bm, t)}

{\it Introduction.}---The field of spintronics is concerned with electric control of spin currents \cite{wolf2001}. For the description of experimentally relevant systems it has, until very recently, been sufficient to consider their coupled spin-charge dynamics in the diffusive regime where the time scale for electron momentum scattering is fast compared to other time scales. The celebrated Valet-Fert theory for electron spin transport in magnetic multilayers \cite{valet1993} and the Dyakonov-Perel drift-diffusion theory for spin generation by the spin Hall effect \cite{dyakonov1971}, for example, fall within this paradigm. 

Very recent experimental developments have brought about solid-state systems, such as ultra-clean encapsulated graphene, in which the momentum scattering time can be much longer than the time scale for electron-electron interactions  \cite{bandurin2016,crossno2016, moll2016,krishna2017}. In this so-called hydrodynamic regime, the electron momentum needs to be included as a hydrodynamic variable and the viscosity of the electron system cannot be neglected \cite{gurzhi1963,black1980,yu1984,molenkamp1995,dyakonov1993,chow1996,spivak2011,torre2015,lucas2015,levitov2016}. The finite electron viscosity leads to several physical consequences, such as a negative nonlocal resistance \cite{bandurin2016} and super-ballistic transport through point contacts \cite{guo2017,krishna2017}. These developments have spurred on a great deal of research, including proposals for measuring the Hall viscosity \cite{scaffidi2017,pellegrino2017,delacretz2017} and connections to strong-coupling predictions from string theory \cite{link2017}.

In a seemingly unrelated development, spin-hydrodynamic generation, i.e.~the generation of voltages from vorticity, was recently experimentally observed in liquid Hg  \cite{takahashi2016}. Spin-hydrodynamic generation is believed to be a consequence of spin-vorticity coupling. Phenomenological theories of spin-vorticity coupling were developed early on \cite{degrootbook} and have been applied to fluids consisting of particles with internal angular momentum such as ferrofluids \cite{felderhof2011}, molecular nanofluids \cite{hansen2009}, and nematic liquid crystals \cite{chueng2002}. In these phenomenological theories, the coupling between orbital angular momentum, i.e.~vorticity of the fluid, and internal angular is governed by a dissipative coefficient, the so-called ``rotational viscosity''. This type of viscosity has been estimated microscopically for classical systems (see e.g. \cite{chueng2002}) and Hg \cite{takahashi2016}, but not for viscous electrons in a crystal.

Motivated by the recent realization of solid-state systems hosting viscous electron fluids, we develop in this Letter the theory for spin-vorticity coupling in such systems. We derive the phenomenological equations describing coupled spin and momentum diffusion, and compute the rotational viscosity microscopically. We apply our theory to viscous electron flow through a point contact and show that the spin densities generated hydrodynamically can be much larger than the ones that are generated by the spin Hall effect in the diffusive transport regime. Our results may therefore stimulate experimental research towards novel ways of spin detection and generation.

{\it Phenomenology.}---We consider two-dimensional (2D) electron systems with approximate translation invariance and approximate rotation invariance around the axis perpendicular to the plane (chosen to be the $\hat{\bm z}$-direction). The conserved quantities of this system are energy, charge, linear momentum in the plane and angular momentum in the $\hat{\bm z}$-direction. For brevity, we do not consider energy conservation explicitly and focus on momentum and angular momentum conservation. In the following, we follow the discussion of Ref.~\cite{degrootbook} and generalize it to include spin diffusion and lack of Galilean invariance. The momentum density is denoted by $\bp (\br,t)$ and is a 2D vector $\bp = (p_x,p_y)$ in the $\hat{\bm x}$-$\hat{\bm y}$-plane with $\br=(x,y)=(r_x,r_y)$. The total angular momentum density in the $\hat{\bm z}$-direction is the sum of orbital angular momentum density $\epsilon_{\alpha\beta}r_\alpha p_\beta$ and spin density $s (\br,t)$ (in the $\hat{\bm z}$-direction). Here, $\epsilon_{\alpha\beta}$ is the 2D Levi-Civita tensor and summation over repeated indices $\alpha,\beta,\gamma,\delta \in \{x,y\}$ is implied. We denote with $\bv$ the conjugate variable to the momentum density, i.e., the velocity, whereas the spin chemical potential, commonly referred to as spin accumulation, $\mu_{\rm s}$ is the conjugate variable to the spin density.

Conservation of linear momentum yields
\begin{equation}
\label{eq:consmomentum}
  \frac{\partial p_\alpha (\br,t)}{\partial t} = - \frac{\partial \Pi_{\alpha \beta} (\br,t)}{\partial r_\beta}~,
\end{equation}
with $\Pi_{\alpha\beta} (\br,t)$ the stress tensor. 
Conservation of angular momentum in the $z$-direction is expressed as
\begin{eqnarray}
\label{eq:consangmomentum}
  \frac{\partial \left[ \epsilon_{\alpha\beta}r_\alpha p_\beta (\br,t) + s (\br,t)\right]}{\partial t} = -\frac{\partial j^J_\alpha(\br,t)}{\partial r_\alpha}~,
\end{eqnarray}
with $j^{J}_{\alpha}(\br,t)$ the $\alpha$-th component of the angular momentum current and in the above equations the summation is over both $\alpha$ and $\beta$. The equation for the spin density is found by subtracting the cross-product of $\br$ with Eq.~(\ref{eq:consmomentum}) from Eq.~(\ref{eq:consangmomentum}) and yields
\begin{equation}
\label{eq:spinbalance}
 \frac{\partial s (\br,t)}{\partial t} = - \frac{\partial j^{\rm s}_\alpha (\br,t)}{\partial r_\alpha} - 2 \Pi^{\rm a} (\br,t)~,
\end{equation}
with $\Pi^{\rm a} (\br,t) = \epsilon_{\alpha\beta} \Pi_{\beta \alpha} (\br,t)/2$ the antisymmetric part of the stress tensor and $j^{\rm s}_\alpha (\br,t) = j^J_\alpha (\br,t)- \epsilon_{\beta\gamma} r_\beta \Pi_{\gamma\alpha} (\br,t)$ the spin current.

A nonzero velocity and spin density increase the energy of the system. By symmetry, a nonzero velocity leads to a contribution $\rho_{\rm kin} \bv^2/2$ to the energy density. This expression defines the kinetic mass density $\rho_{\rm kin}$, such that $\bp (\br,t) = \rho_{\rm kin} \bv (\br,t)$ \cite{deboer2017}. For the case that is of interest to us, i.e., 2D electrons with spin-orbit coupling, the kinetic mass density is not equal to the average mass density $\rho$ because spin-orbit coupling breaks Galilean invariance. Likewise, a nonzero spin density contributes $\chi_{\rm s} \mu_{\rm s}^2/2$ to the energy density, where $\chi_{\rm s}$ is the static spin susceptibility, so that $s (\br,t) = \hbar \chi_{\rm s} \mu_{\rm s} (\br,t)$. These terms in the energy density lead to contributions to the entropy production from which relations between the fluxes (the spin current and antisymmetric part of the pressure tensor) and the forces (spin accumulation and velocity) are derived phenomenologically. In terms of $\mu_{\rm s} (\br,t)$ and $\bv (\br,t)$ we have for the antisymmetric part of the pressure tensor that \cite{degrootbook}
\begin{equation}
\label{eq:aspartstressenergytensor}
\Pi^{\rm a} (\br,t) = - \eta_{\rm r} \left[ \omega (\br,t)- 2\mu_{\rm s}(\br,t)/\hbar \right]~,
\end{equation}
with $\omega (\br,t) = \epsilon_{\alpha\beta} \partial v_\beta(\br,t)/\partial r_\alpha$ the vorticity and $\eta_{\rm r}$ the {\it rotational viscosity}. The above expression shows that angular momentum is transferred, by spin-orbit coupling, between orbital and spin degrees of freedom until the antisymmetric part of the pressure tensor is zero. For the spin current we have that $j^{\rm s}_\alpha (\br,t) = -\sigma_{\rm s}  \partial \mu_{\rm s} (\br,t)/\partial  r_\alpha = -D_{\rm s} \partial s (\br,t)/\partial r_\alpha$ which defines the spin diffusion constant $D_{\rm s}$ and spin conductivity $\sigma_{\rm s}$, which obey the Einstein relation $\sigma_{\rm s} = \hbar D_{\rm s} \chi_{\rm s}$.  Note that we are omitting an advective contribution $\sim v_\alpha s$ to the spin current as we restrict ourselves to the linear-response regime. Inserting these results for the fluxes into Eq.~(\ref{eq:spinbalance}) and using Eq.~(\ref{eq:consmomentum}) leads to
\begin{eqnarray}
\label{eq:finalhydroeqs}
 && \frac{\partial s (\br,t)}{\partial t} = D_{\rm s} \nabla^2 s (\br,t) \nonumber \\
&&  ~+ 
2 \eta_{\rm r} \left[ \omega (\br,t) - \frac{2 s(\br,t)}{\hbar^2 \chi_{\rm s}}\right] - \frac{s (\br,t)}{\tau_{\rm sr}}~; \nonumber \\
&& \rho_{\rm kin} \frac{\partial v_\alpha (\br,t)}{\partial t} = -\frac{e \rho E_\alpha}{m} + \nu \rho_{\rm kin}\nabla^2 v_\alpha (\br,t) \nonumber \\ 
&& ~+ \eta_{\rm r} \epsilon_{\alpha\beta} \frac{\partial}{\partial r_\beta}  \left[ \omega (\br,t) - \frac{2 s(\br,t)}{\hbar^2 \chi_{\rm s}}\right] - \frac{\rho_{\rm kin} v_\alpha (\br,t)}{\tau_{\rm mr}}.
\end{eqnarray}
In the above we have assumed the linear-response regime and introduced the kinematic viscosity $\nu$ using that the symmetric part of the stress tensor is given by $\Pi_{\alpha\beta}=\nu \rho_{\rm kin} \partial v_\alpha/\partial r_\beta$. Furthermore, we have added spin and momentum relaxation terms, parameterized by the phenomenological time scales $\tau_{\rm sr}$ and $\tau_{\rm mr}$, respectively. We have also included an electric field ${\bm E}$ (the electron has charge $-e$). 

Eqs.~(\ref{eq:finalhydroeqs}) are the main phenomenological equations for spin density and velocity. The term proportional to $\eta_{\rm r}$ in the first equation describes generation of spin accumulation in response to vorticity, e.g., spin-vorticity coupling. 

In the steady state the hydrodynamic equations are characterized by three length scales. The first is a length scale that results from the spin-vorticity coupling equal to $\ell_{\rm sv} = \sqrt{D_{\rm s} \hbar^2 \chi_{\rm s}/(2 \eta_{\rm r})}$, which is the characteristic length over which the orbital and spin angular momentum equilibrate. Furthermore, we have the spin diffusion length $\ell_{\rm sr} = \sqrt{D_{\rm s} \tau_{\rm sr}}$ that determines the length scales for relaxation of spin due to impurities, and the momentum diffusion length $\ell_{\rm mr} = \sqrt{\nu \tau_{\rm mr}}$. The most interesting regime, which occurs in the limit of strong spin-orbit coupling relative to momentum and spin relaxation, is the one where $\ell_{\rm sv}$ is the shortest length scale. In this case the spin density locally follows the vorticity, which is determined by the electron flow. 

{\it Application.}---We consider electron flow through a point contact (PC) \cite{krishna2017,guo2017} driven by a voltage $V$. Taking $\tau_{\rm mr}, \tau_{\rm sr} \to \infty$ we have from Ref.~\cite{guo2017} for the velocity distribution at the PC that
\begin{equation}
\label{eq:velocitypointcontact}
  v_y (x) = -\frac{\pi \rho e V}{4 m \nu \rho_{\rm kin}} \sqrt{\left(\frac{w}{2}\right)^2-x^2}~,
\end{equation}
where the flow is in the $y$-direction and $w$ is the PC width. From Eq.~(\ref{eq:finalhydroeqs}), in the limit $\ell_{\rm sv} \ll w$ the steady-state spin density generated at the PC by spin-vorticity coupling in the hydrodynamic regime is then
\begin{equation}
\label{eq:spindensitypointcontact}
\frac{s (x)}{\hbar^2 \chi_{\rm s} j^{\rm c}} = -\frac{m}{\pi e w \rho}  \frac{4 x}{\sqrt{(w/2)^2-x^2}}~,
\end{equation}
where $j^{\rm c} = -e \rho\int dx\, v_y (x)/(m w)$ is the average current density.

Let us compare Eq.~(\ref{eq:spindensitypointcontact}) with the spin density generated by the spin Hall effect in the diffusive limit. In the latter case, the spin accumulation is determined by $\partial^2 \mu_{\rm s}/\partial x^2 = \mu_{\rm s}/\ell_{\rm sr}^2$, which follows from Eqs.~(\ref{eq:finalhydroeqs}) in the limit  $\ell_{\rm sr} \ll \ell_{\rm sv}$, together with the expression $j^{\rm s}_y=-\sigma_{\rm s} \partial \mu_x/\partial x +\theta_{\rm SH} \hbar j^{\rm c}_y/(2e)$ for the spin current. Here $j^{\rm c}_y = \sigma_{\rm e} E_y$ is the diffusive charge current through the PC, with $\sigma_{\rm e} =e^2 \rho^2 \tau_{\rm mr}/(m^2 \rho_{\rm kin})$ the electrical conductivity and $\theta_{\rm SH}$ the spin Hall angle. Using the boundary conditions $j^{\rm s}(-w/2) = j^{\rm s}(w/2)=0$, we find for the spin density in the diffusive limit that
\begin{equation}
\label{eq:sdiff}
 \frac{s_{\rm diff} (x)}{\hbar^2 \chi_{\rm s} j^{\rm c}_y} = \frac{\theta_{\rm SH} \ell_{\rm sr}}{2e\sigma_{\rm s}}  {\rm sech} \left(\frac{w}{2\ell_{\rm sr}}\right) {\rm sinh} \left( \frac{x}{\ell_{\rm sr}}\right)~.
\end{equation}
A crucial difference is thus that for diffusive spin transport and when $w \gg \ell_{\rm sr}$, the spin density is only nonzero within a distance $\sim \ell_{\rm sr}$ away from the edges of the PC, while when $w \gg \ell_{\rm sv}$ and in the hydrodynamic limit, the spin density [see Eq.~(\ref{eq:spindensitypointcontact})] is nonzero everywhere (except at $x=0$ where it vanishes by symmetry). 

In both hydrodynamic and diffusive limits, the maximum spin density occurs at the edges. In the hydrodynamic limit the spin density formally diverges as $|x| \to w/2$, since the vorticity that results from the velocity in Eq.~(\ref{eq:velocitypointcontact}) diverges in the same limit. This divergence is, however, unphysical, as there will be a microscopic length scale $\ell_{\rm edge}$ over which the velocity goes to zero near the edge of the sample, resulting in a maximum spin density of $|s(\pm w/2)|/(\hbar^2 \chi_{\rm s} j^{\rm c}) \sim  m/(e \rho \ell_{\rm edge})$ near the edges of the sample. We expect the latter to be much larger than the maximum spin density $|s_{\rm diff} (\pm w/2)/ (\hbar^2 \chi_{\rm s} j^{\rm c}) |\sim m^2 \theta_{\rm SH} \ell_{\rm sr} / (e \hbar \rho \tau_{\rm mr})$ generated by the spin Hall effect in the diffusive regime (where we estimated $\sigma_{\rm s} \sim \hbar \rho \tau_{\rm mr}/m^2$), because $\hbar \tau_{\rm mr}/(m \theta_{\rm SH} \ell_{\rm sr}) \sim \ell_{\rm mr}/(\theta_{\rm SH} k_{\rm F} \ell_{\rm sr})$ is expected to be much larger than the microscopic length scale $\ell_{\rm edge}$. Here, $k_{\rm F}$ is the Fermi wave number. 
 
{\it Microscopic theory.}---We proceed by calculating the rotational viscosity microscopically. This is most easily achieved \cite{footnote} by noting that even when spin relaxation due to impurities is absent ($\tau_{\rm sr} \to \infty$), the spin-vorticity coupling opens a channel for spin relaxation, with rate $4\eta_{\rm r}/\hbar^2 \chi_{\rm s}$, which microscopically stems from the combined effect of spin-orbit coupling and electron-electron interactions. Hence, $\eta_{\rm r}$ can be extracted from the retarded spin-spin response function (for spin in the $\hat{\bm z}$-direction) at zero wave vector, denoted by $\chi^{(+)}_{\rm s}(\omega)$, when this response function is computed for a clean system with spin-orbit coupling and interactions. From Eqs.~(\ref{eq:finalhydroeqs}) we find that for ${\bm v} = {\bm 0}$  this response function has the form
\begin{equation}
\label{eq:spinresponsefct}
\chi^{(+)}_{\rm s}(\omega) = \frac{\chi_{\rm s}}{1- i \omega \hbar^2\chi_{\rm s}/(4 \eta_{\rm r})}~.
\end{equation}
Hence, we have that
\begin{equation}\label{eq:microexpressionrotvisc} 
 \frac{1}{\eta_{\rm r}} = - \left(\frac{2}{\hbar \chi_{\rm s}}\right)^2 \lim_{\omega \to 0} \frac{{\rm Im}[\chi^{(+)}_{\rm s}(\omega)]}{\omega}~.
\end{equation}

As a representative example, we compute the rotational viscosity using standard linear-response techniques for a 2D electron gas with Rashba spin-orbit coupling, which has the following Hamiltonian \cite{manchon2015}:
\begin{equation} 
 \hat{\cal H} = \int d\br \sum_{\sigma \in \{ \uparrow, \downarrow\}}\hat \psi^\dagger_\sigma (\br) \left[ - \frac{\hbar^2 \nabla^2}{2m} + \lambda \hbar \hat{\bm z} \cdot \left(\frac{\nabla}{i} \times \bm{\tau}\right) \right]\hat \psi_\sigma (\br)~,
\end{equation}
where $\hat \psi_\sigma (\br)$ [$\hat \psi_\sigma^\dagger (\br)$] is an electron annihilation [creation] operator 
and $\bm{\tau}$ is a vector of Pauli matrices. The unit vector in the $\hat{\bm z}$-direction is denoted by $\hat{\bm z}$. The constant $\lambda$ parametrizes the strength of spin-orbit interactions. The spin density operator in imaginary time $\tau$ is $\hat s(\br,\tau) = \hbar[ \hat \psi_\uparrow^\dagger (\br,\tau) \hat \psi_\uparrow (\br,\tau) - \hat \psi_\downarrow^\dagger (\br,\tau) \hat \psi_\downarrow (\br,\tau)]/2$, where the dependence on $\tau$ of the electron creation and annihilation operators indicates their corresponding Heisenberg evolution in imaginary time. 
We have for the imaginary-time spin-spin response function
\begin{equation}\label{eq:spin_bubble}
\chi_{\rm s}(i \omega_n) = \frac{1}{\hbar} \int d\br \int_0^{\hbar\beta} d\tau\langle \hat s (\br,\tau) \hat s (\br,0) \rangle_0 e^{i \omega_n \tau}~,
\end{equation}
where $i \omega_n = 2 \pi n/(\hbar\beta)$ is a bosonic Matsubara frequency with $\beta=1/(k_{\rm B} T)$ the inverse thermal energy, and the expectation value $\langle \cdots \rangle_0$ is taken at equilibrium. Neglecting vertex corrections due to interactions, this is worked out to yield
\begin{eqnarray}
\label{eq:chiimagbubble}
 \chi_{\rm s} (i \omega_n) &=& -\frac{1}{4 \hbar V} \sum_{\bk} \sum_{\delta \neq\delta'} \int d \hbar\omega d \hbar\omega' A_{\delta} (k,\omega) A_{\delta'} (k,\omega') \nonumber \\
 &&\times \left[\frac{N(\hbar \omega) - N(\hbar \omega')}{ \omega-\omega' +i  \omega_n}\right]~,
\end{eqnarray}
with $N(\hbar \omega) = \left[e^{\beta\left(\hbar \omega - \mu \right)}+1\right]^{-1}$ the Fermi-Dirac distribution function at chemical potential $\mu$. The spectral functions $A_{\delta} (k,\omega)$ are labeled by the Rashba spin-orbit-split band index $\delta = \pm$. We incorporate electron-electron interactions into the spectral function by taking them equal to Lorentzians broadened by the electron collision time $\tau_{\rm ee}$ [this corresponds to dressing bare propagator lines in the spin bubble in Eq.~(\ref{eq:spin_bubble}) by self-energy insertions], i.e., 
\begin{equation}\label{eq:spectralfct}
 A_\delta (k,\omega) = \frac{\hbar}{2\pi \tau_{\rm ee}} \frac{1}{\left[\hbar \omega-\hbar \omega_\delta (k)\right]^2+\left(\frac{\hbar}{2\tau_{\rm ee}}\right)^2}~,
\end{equation}
where $\hbar \omega_\delta (k) = \hbar^2 k^2/2m +\delta \hbar \lambda k$ is the Rashba band dispersion. Inserting Eq.~(\ref{eq:spectralfct}) into Eq.~(\ref{eq:chiimagbubble}) and performing a Wick rotation $i \omega_n \to \omega+i 0^{+}$ yields
\begin{equation}\label{eq:rotviscstrongSOC}
\eta_{\rm r} = \frac{4\pi^2 \hbar^4 \chi^2_{\rm s}}{m\tau_{\rm ee}} \left[  2\pi + \frac{8 \left(\frac{ \mu \tau_{\rm ee}}{\hbar}\right)}{1+4\left(\frac{ \mu \tau_{\rm ee}}{\hbar}\right)^2} + 4\tan^{-1} \left(\frac{2 \mu \tau_{\rm ee}}{\hbar}\right) \right]~,
\end{equation}
where we took $\lambda \to 0$. In the limit $\mu \tau_{\rm ee}/\hbar \gg 1$, we have $\eta_{\rm r} = \pi \hbar^4 \chi^2_{\rm s}/(m\tau_{\rm ee})$.

Since we have neglected vertex corrections, the result in Eq.~(\ref{eq:rotviscstrongSOC}) does not vanish in the $\lambda \to 0$ limit and is strictly speaking only valid when spin-orbit coupling is so strong that the spin-vorticity coupling is limited by electron-electron interactions, i.e., when $\lambda  k_{\rm F} \tau_{\rm ee}\gg 1$. In the opposite limit, where the bottleneck for spin relaxation is the spin-orbit coupling, we perform a Fermi's Golden Rule calculation to determine the decay rate of a spin polarization to second order in the strength of the spin-orbit interactions.  This gives at low temperatures that
\begin{eqnarray}
\label{eq:spinrotweakbeforeint}
\eta_{\rm r}=- \frac{\pi\hbar}{8 } \int \frac{d\bk}{(2\pi)^2}
A^2 (k,\mu) (\lambda \hbar k)^2~,
\end{eqnarray}
where $A(k,\mu)$ is the spectral function obtained from Eq.~(\ref{eq:spectralfct}) by replacing $\hbar \omega_\delta (k) \to \hbar^2 k^2/2m$. Carrying out the remaining integral gives
\begin{equation}\label{eq:rotviscweakSOC}
\eta_{\rm r} = \frac{m \lambda^2}{2\hbar}\left[1+\pi\left(\frac{\mu\tau_{\rm ee}}{\hbar}\right) +2\left(\frac{\mu\tau_{\rm ee}}{\hbar} \right)\tan^{-1} \left(\frac{2 \mu \tau_{\rm ee}}{\hbar}\right)\right]~,
\end{equation} 
which indeed vanishes as $\lambda \to 0$. When $\mu \tau_{\rm ee}/\hbar \gg 1$, we have that 
$ \hbar \eta_{\rm r} \sim (\lambda k_{\rm F}) (\lambda k_{\rm F} \tau_{\rm ee})$, showing the dependence on the small parameter $\lambda k_{\rm F} \tau_{\rm ee} \ll 1$ explicitly. Interestingly, since the kinematic viscosity $\nu \propto \tau_{\rm ee}$, we have that the rotational viscosity $\eta_{\rm r} \propto 1/\nu$ in the limit of strong spin-orbit coupling and $\eta_{\rm r} \propto \nu$ in the limit of weak spin-orbit coupling, with a maximum rotational viscosity when $\lambda k_{\rm F} \tau_{\rm ee} \sim 1$.

{\it Estimates.}---Next, we estimate the spin-vorticity coupling for graphene with proximity-induced spin-orbit coupling. We take $\lambda \hbar k_{\rm F}$ to be on the order of $1~{\rm meV}$ \cite{wang2015}. Furthermore, we take $\tau_{ee} \sim 100~{\rm fs}$ \cite{bandurin2016}. We thus have that $\lambda \hbar k_{\rm F}$ is about one order of magnitude smaller than $\hbar/\tau_{\rm ee}$ and use the weak spin-orbit coupling expression in Eq.~(\ref{eq:spinrotweakbeforeint}). Evaluating Eq.~(\ref{eq:spinrotweakbeforeint}) for a linear dispersion $\hbar v_{\rm F} k$, where $v_{\rm F} \sim 10^6$ m$/$s is the graphene Fermi velocity, we find that 
\begin{equation}
\label{eq:rotviscweakSOCgraphene}
\eta_{\rm r} \sim \frac{(\lambda \hbar k_{\rm F})^2}{\hbar v^2_{\rm F}} \left(\frac{\mu\tau_{\rm ee}}{\hbar}\right)~,
\end{equation} 
using $\mu \tau_{\rm ee} \gg \hbar$. We estimate the corresponding inverse time scale as
\begin{equation}
\label{eq:rotviscweakSOCgraphene}
\frac{\eta_{\rm r}}{\hbar^2 \chi_{\rm s}} \sim \frac{(\lambda \hbar k_{\rm F})^2}{\hbar^3 \chi_{\rm s} v^2_{\rm F}} \left(\frac{\mu\tau_{\rm ee}}{\hbar}\right) \sim 100~{\rm GHz}~,
\end{equation} 
where we took $\mu \tau_{\rm ee}/\hbar \sim 10$, and estimated the spin susceptibility as $\chi_{\rm s}  \sim D(\mu)$, with the density of states at the Fermi level $D (\mu) \sim \sqrt{n_{\rm e}}/(\hbar v_{\rm F})$, and the electron number density $n_{\rm e} \sim 10^{12}$ cm$^{-2}$ \cite{bandurin2016}. 

To estimate the corresponding length scale $\ell_{\rm sv}$, we assume that spin diffusion is in the hydrodynamic regime determined by electron-electron interactions that lead to spin drag \cite{damico2000}. We then have for the spin diffusion constant that $D_{\rm s} \sim \hbar \rho \tau_{\rm ee}/(m^2 \chi_{\rm s})$. The spin-vorticity length scale is then $\ell_{\rm sv} \sim v_{\rm F} \hbar \sqrt{\tau_{\rm ee} \chi_{\rm s}/\eta_{\rm r}}\sim 1~{\rm \mu m}$. This is the same order of magnitude as the momentum relaxation length scale $\ell_{\rm mr}$ \cite{bandurin2016}, so that the rotational viscosity appears to be high enough to lead to observable spin-vorticity coupling. Moreover, the limit where $\ell_{\rm sv} < \ell_{\rm mr}$ seems to be within experimental reach. Note that in the regime of weak spin-orbit coupling we have  for the spin relaxation the Dyakonov-Perel result that $1/\tau_{\rm sr} \propto \tau_{\rm mr }$ \cite{dyakonov1972}, which yields that in the hydrodynamic regime we have 
$\ell_{\rm sr} \sim \ell_{\rm sv}  \sqrt{\tau_{\rm ee}/\tau_{\rm mr}} \gg \ell_{\rm sv}$.

A simple interpretation of the spin-vorticity coupling is that the electron spins are polarized by an effective magnetic field $\hbar \omega (\br,t)/\mu_{\rm B}$, with $\mu_{\rm B}$ the Bohr magneton, in the frame that rotates with the electron flow vorticity. We estimate the vorticity $\omega \sim v/\ell_{\rm mr}$ using $\ell_{\rm mr}\sim 0.1$-$1~{\rm \mu m}$, and a drift velocity of $v \sim 100~{\rm m}/{\rm s}$ \cite{bandurin2016}, which yields a substantial effective magnetic field of $1$-$10~{\rm mT}$. 

{\it Discussion and conclusions.}---We have developed the theory for spin-vorticity coupling in viscous electron fluids, both phenomenologically and microscopically, and we have estimated that the proximity-induced spin-orbit coupling in graphene is large enough for observable effects. As an example, we predict a large spin polarization induced by spin-hydrodynamic generation in a PC. This large spin density  may e.g. be observed optically \cite{sih2005} or via nitrogen-vacancy centre magnetometry \cite{maze2008, rondin2014}. The imaged spin density would provide a fingerprint of the vorticity of the electron flow. 

An interesting direction for future research is generalization of the phenomenological and microscopic derivation to other spin-orbit couplings, including, in particular, also the effects of violation of translational and rotational invariance beyond the phenomenological relaxation terms that we included here. One example would be that of Weyl semi-metals that naturally have sizeable spin-orbit coupling and have also been reported to be able to reach the hydrodynamic regime \cite{gooth2017}. Other candidates are bismuthene \cite{reis2017} and stanene \cite{zhu2015} that combine strong spin-orbit coupling with high mobility.   Further interesting directions of research include incorporating effects of a magnetic field and computation of the rotational viscosity in the regime where spin-orbit interactions and electron-electron interactions are comparable in magnitude. In this regime, the crossover from weak-to-strong spin-orbit coupling takes place, whereas inclusion of momentum-relaxing scattering would lead to a crossover from the spin-vorticity coupling to the spin Hall effect.  

{\it Acknowledgements.}---We thank Denis Bandurin, Eugene Chudnovsky, and Harold Zandvliet for useful comments. R.D. is member of the
D-ITP consortium, a program of the Netherlands Organisation
for Scientific Research (NWO) that is funded by
the Dutch Ministry of Education, Culture and Science
(OCW). This work is in part funded by the Stichting voor Fundamenteel
Onderzoek der Materie (FOM) and the European Research Council (ERC). M.P. is supported by the European Union's Horizon 2020 research and innovation programme under grant agreement No.~696656 ``GrapheneCore1''.


\begin{thebibliography}{99}
\bibitem{footnote} Alternatively, the rotational viscosity may be evaluated using the linear-response relation in Eq.~(\ref{eq:aspartstressenergytensor}). From this relation we find that 
	$\eta_{\rm r}$ can be extracted from the correlation function $K (\br-\br';\tau-\tau') =\langle \hat \Pi^{\rm a} (\br,\tau) \hat s (\br',\tau')\rangle_0$ at zero wave vector so that
	$\eta_{\rm r} = K(k \to 0,i\omega_n)/2$, where the factor $1/2$ compensates for the factor $2$ in Eq.~(\ref{eq:aspartstressenergytensor}).  Using Eq.~(\ref{eq:spinbalance}) at zero wave vector we have that $K(k\to 0;i\omega_n) = -  i \hbar^2 \omega_n \chi_{\rm s}(i\omega_n)/2$, so that insertion of Eq.~(\ref{eq:spinresponsefct}) indeed yields that $ K(k \to 0,i\omega_n) = 2 \eta_{\rm r}$ in the ``fast" $\omega_n \gg \eta_{\rm r} /(\hbar^2 \chi_{\rm s})$ limit \cite{luttinger1964}. For completeness, we also compute the kinetic mass density for the Rashba model in the noninteracting limit. We then have that the momentum density is given by
\[	
	\bp  =\sum_\delta \int \frac{d\bk}{(2\pi)^2} \hbar \bk A_\delta (k,\omega) N (\hbar \omega -\mu -\hbar \bk \cdot \bv)~,
\]
	which yields $\bp = \rho_{\rm kin} \bv$ for $\bv \to {\bm 0}$, with the kinetic mass density $\rho_{\rm kin} = \rho [1 + \lambda^2 m/2\mu + {\mathcal O}(\lambda^4)]$. The correction, i.e.~the second term between the brackets, is typically small since one almost always has that $ \lambda^2 m/2\mu \ll 1$.
\end{thebibliography}
\end{document}